# Multi-wavelength ghost imaging: a review

Tong Tian[1,2,*], Sukyoon Oh[1,2,*], Christian Spielmann[1,2,*]

1 Institute of Optics and Quantum Electronics, Abbe Center of Photonics, Friedrich-Schiller-University Jena, Max-Wien-Platz 1, 07743, Jena, Germany

2 Helmholtz Institute Jena, Fröbelstieg 3, 07743 Jena, Germany

*Corresponding authors:  christian.spielmann@uni-jena.de

**Abstract:** Ghost imaging (GI) forms images from intensity-correlation data collected by a single-pixel detector, decoupling illumination and sensing. Since its quantum-photon origins, the technique has evolved through classical pseudothermal, computational and deep-learning variants to span an unprecedented spectral range—from extreme-ultraviolet (XUV) to terahertz (THz) waves and even matter waves. This review traces that evolution, highlighting how wavelength dictates modulators, detectors and propagation physics and, in turn, the attainable penetration depth, resolution and dose. We survey X-ray/XUV implementations that deliver low-damage microscopy, visible/near-IR systems that achieve video-rate lidar through fog and water, mid-IR platforms that extract molecular fingerprints in photon-starved conditions, and THz schemes that provide non-destructive inspection of concealed structures. These developments position multi-wavelength GI as a powerful, low-dose imaging alternative where conventional focal-plane arrays face limitations.

## 1. Introduction

Ghost imaging (GI) can be carried out across an enormous span of the electromagnetic spectrum, and the choice of wavelength strongly shapes the accessible hardware, image quality, penetration depth and the sorts of samples[1]. In a typical GI experiment, two correlated optical fields are generated. The test beam interrogates the object and is collected by a single-pixel (bucket) detector, whereas the reference beam is measured with spatial or temporal resolution but never interacts with the sample. An image is reconstructed by computing the second-order correlation between these beams, enabling imaging even at wavelengths where conventional cameras are ineffective [2][3]. This decoupling of illumination and detection makes GI especially valuable in spectral regions where cameras, modulators, or high-speed detectors are limited or noisy [4][5].

The concept of quantum entanglement GI was proposed by Klyshko in 1988[6]. Pittman and others verified it experimentally in 1995, based on entangled photon pairs, which Pittman interpreted as the result of coherent superposition of entangled photon pairs[7]. Subsequently, researchers began to shift their focus to the field of classical light sources, such as incoherent thermal light sources[8], particularly rubidium hollow-cathode lamps[9][10], polarized laser beams[11], and laser beams through rotating diffusers[12][13]. The advent of computational GI (CGI) in 2008[14] marked a pivotal shift, demonstrating GI's potential with classical optics and leading to the development of methods such as differential GI (DGI)[15], compressive sensing GI (CSGI)[16], and deep learning ghost imaging (DLGI)[17]. These advancements have expanded GI's application to challenging imaging contexts like multiwavelength and scattering media imaging. Notably, GI leverages the flexibility of single-pixel detectors to offer distinct advantages over traditional array detectors, particularly in terms of adaptability and sensitivity across different wavelengths. Unlike multi-pixel array detectors such as CMOS and CCD cameras that are optimized for specific wavelength ranges and can suffer from quantum efficiency degradation outside these ranges, single-pixel detectors are inherently more versatile. They can be designed or selected to have high sensitivity across a broader spectrum of wavelengths, including extreme ultraviolet (XUV/EUV) and other ranges where traditional sensors typically underperform or are unavailable. Additionally, the simplicity and compactness of single-pixel setups allow for easier integration into existing systems and can lead to reductions in both cost and complexity of imaging setups.

This paper presents a comprehensive overview of correlation-based ghost imaging, with a focus on its implementation across multiple wavelengths. We trace the technique's evolution from the earliest demonstrations using particle correlations, through X-ray and extreme-ultraviolet (XUV) ghost imaging, to developments in the visible band, infrared regimes, and, most recently, terahertz wavelengths. Along the way we analyze the enabling hardware and computational strategies that drive high-fidelity image reconstruction across these disparate spectral windows. Finally, we assess how emerging artificial-intelligence methods, ranging from compressed-sensing priors to large-scale generative models, are poised to accelerate multi-wavelength ghost imaging and outline the key challenges and opportunities that will shape the field in the coming decade.

**2. Theory**

2.1 Principle

Quantum entanglement-based ghost imaging (GI) is fundamentally limited by low efficiency and practical challenges [1]. The introduction of classical light sources marked a significant advance in understanding the underlying principles of GI. Among these, pseudothermal light gained prominence due to its simplicity, accessibility, and extensive use in both experimental and computational ghost imaging (CGI) studies. The basic structure of a CGI setup is illustrated in Figure 1.The light beam emitted by the light source is divided into two paths through a beam splitter: one path is irradiated to the target object, and the light reflected or transmitted by the object is received by a bucket detector without spatial resolution capability to record the total light intensity information; the other path is directly incident on a charge-coupled device with spatial resolution capability to record the spatial distribution of the reference light field. By performing correlation calculations on these two light signals, the image information of the object can be reconstructed. The correlation operation can be expressed as follows[2]:

$$T(x,y) = <S_i \cdot I_i(x,y)> - <S> \cdot <I(x,y)> \qquad (1)$$

where *T(x,y)* represents the image value reconstructed at position *(x,y)*, $S_i$ is the total light intensity received by the bucket detector in the i-th measurement, $I_i(x,y)$ is the intensity distribution of the i-th reference light field at that position, and $<\cdot>$ indicates the average of all measurements.

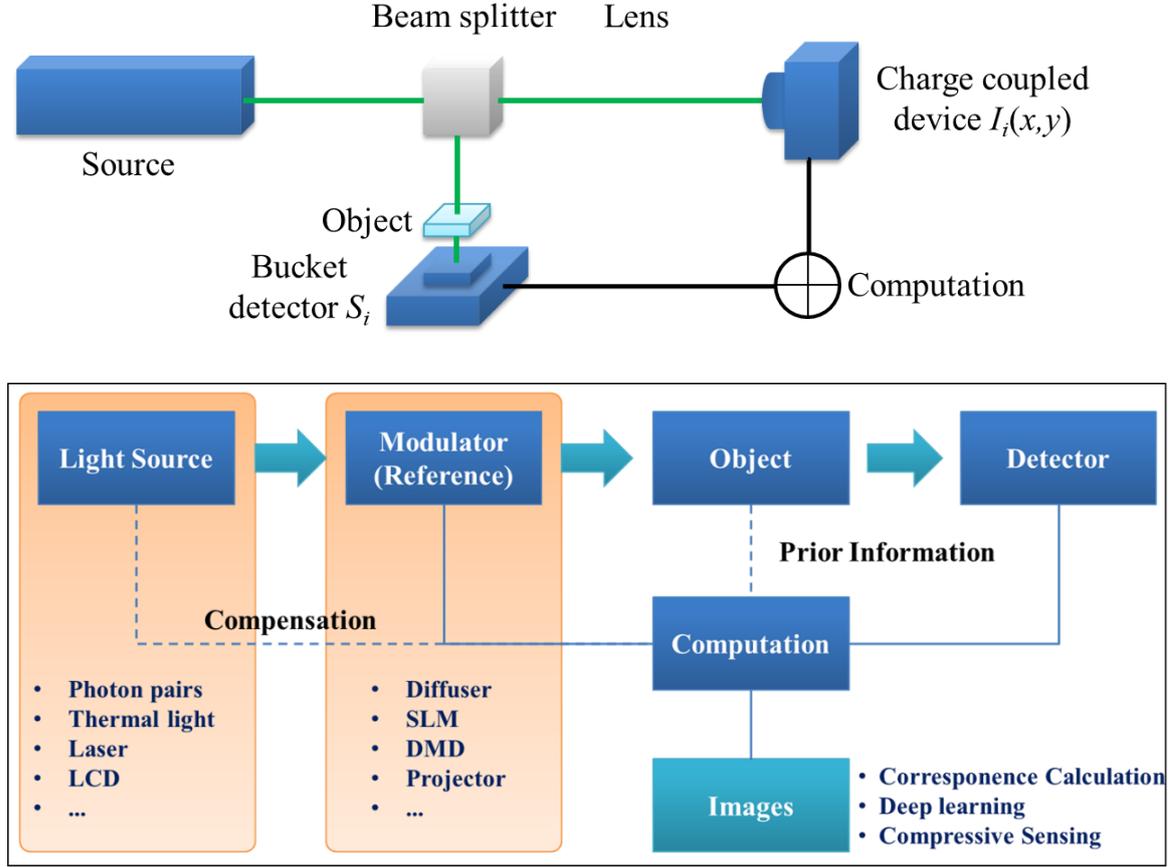

Fig. 1 Experimental setup of ghost imaging and the components and workflow of GI system using computational techniques for image reconstruction. The physical composition of a GI system can be generally divided into four parts: Light source, modulator or reference, object to be imaged and the single-pixel detector. All of the four parts can be connected by computational means. In the modulator section, SLM stands for spatial light modulator and DMD stands for digital micromirror device.

To enhance the imaging quality of GI, differential ghost imaging (DGI) was introduced.[15] In DGI, the total intensity of light source is considered and used as a deviation weight, making improvements when imaging objects with high transparency or reflection, thereby effectively suppressing the background noise and improving the signal-to-noise ratio. The differential correlation calculation method can be expressed as[15]:

$$T_{DGI}(x,y) = <(S_i - \frac{<S>}{<R>}R_i)\cdot(I_i(x,y)-<I(x,y)>)> \quad (2)$$

Compared with Eq. (1), the total intensity of the reference beam *R* is taken into consideration in DGI depicted in Eq. (2). This method enhances the relative fluctuation value of the object information through the differential signal, which is especially suitable for highly transmissive objects. It can obtain higher imaging quality at a lower sampling rate, while reducing the influence of external noise and system sensitivity on the reconstruction results.

To constrain imaging noises caused by light source fluctuations, normalized ghost imaging was also proposed[18].

$$T_{NGI}(x,y) =< (\frac{S_i}{<S>} - 1) \cdot (\frac{I_i(x,y)}{<I(x,y)>} - 1) > \tag{3}$$

Normalization processing standardizes the intensity distribution of the reference light field, making the imaging process insensitive to changes in the absolute value of light intensity, thereby improving the stability and robustness of imaging. Researchers also normalized ghost images by second-order coherence[19].

In 2008, compressive sensing (CS) was applied to single-pixel cameras [14]. CS leverages the sparsity of a signal—that is, the assumption that the signal contains only a few non-zero elements in a specific domain. It enables efficient sampling and reconstruction by acquiring a small number of random measurements, significantly below the Nyquist rate, and then reconstructing the signal using nonlinear algorithms. The mathematical expression of compressed sensing is[20]:

$$y = \Phi x + n \tag{5}$$

where $y$ is the measurement vector consisting of $M$ compressed samples obtained via random linear projections of the original signal $x$ through the measurement matrix $\Phi$. The term $n$ represents measurement noise inherent in analog systems. The signal $x$ can be sparsely represented under a certain transformation basis $\Psi$ as

$$x = \Psi\alpha \tag{6}$$

where, $\Psi$ is a basis matrix and $\alpha$ is a sparse coefficient vector. To reconstruct $x$ from the compressed measurements, the following optimization problem is typically solved:

$$min||\alpha||_1 \; s.t. ||y - \Phi\Psi\alpha||_2 \leq \varepsilon \tag{7}$$

Here, $s.t$ is the abbreviation of *subject to*, which means "under the condition of...". $||\alpha||_1$ is the $L_1$ norm of the sparse vector $\alpha$, which helps to find sparse solutions. $\varepsilon$ corresponds to the noise energy level. In the absence of noise, the constraint can be simplified to $y = \Phi\Psi\alpha$. Compressive sensing has also been successfully applied in entangled-photon ghost imaging [21], and has enabled super-resolution ghost imaging [22].

In recent years, the rapid progress of deep learning has yielded state-of-the-art results in image enhancement, recognition, and segmentation, inspiring researchers to translate these gains to optical imaging. In 2017, Lyu et al.[17] were the first to embed deep learning into computational correlation imaging. They reconstructed a series of noisy, low-sampling-rate images with a traditional ghost-imaging algorithm, paired each result with its high-fidelity ground truth, and used the resulting data set to train a neural network that denoises and refines the low-rate reconstructions. Building on this idea, He et al.[23] designed a dedicated GI Convolutional Neural Network and, through both simulations and experiments, demonstrated substantial gains in reconstruction quality under aggressive undersampling. That same year, Shimobaba et al.[24] independently confirmed the benefits of deep learning for ghost imaging in a series of proof-of-concept experiments. Since these milestones, deep neural networks have become integral to computational correlation imaging[25], not only improving reconstruction fidelity but also opening the door to multi-wavelength ghost imaging.

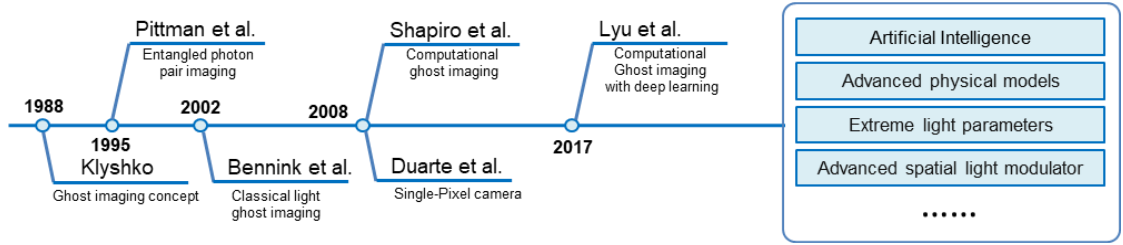

Fig. 2 Timeline of developments of ghost imaging. Originating from entangled photon pair imaging in 1988, ghost imaging has undergone significant advancements. Current trends indicate a shift toward increased intelligence and broader applicability across various fields.

## 2.1 Evaluation Metrics for Ghost Imaging Quality

The assessment of imaging quality in GI relies on several pivotal metrics: root mean squared error (RMSE), peak signal-to-noise ratio (PSNR), contrast-to-noise ratio (CNR), structural similarity (SSIM) and spatial resolution, and so on. RMSE is expressed as [26]:

$$RMSE = \sqrt{\frac{1}{N}\sum_{i=1}^{N}(y_i - \hat{y}_i)^2} \qquad (8)$$

Where $N$ is total number of pixels. Furthermore, $y_i$ and $\hat{y}_i$ are the actual value and ground-truth value of each pixel, respectively. RMSE quantifies reconstruction fidelity by measuring noise suppression and structural preservation.

PSNR is defined as [26]

$$PSNR = 10\lg\left(\frac{MAX_I^2}{MSE}\right) \qquad (9)$$

where MSE is the mean squared error between the reconstructed and ground-truth images, which is the square of RMSE shown in eq.(8). The $MAX_I$ in eq. (9) is the maximum value of all the pixels.

CNR is expressed as[27][28]

$$CNR = \frac{S_{obj} - S_{bkg}}{\sigma_{img}} \qquad (10)$$

where $S_{obj}$ and $S_{bkg}$ are the mean intensities of the object region and background region, respectively. Moreover, $\sigma_{img}$ is the standard derivations of the image. CNR evaluates the distinguishability of target features against background noise, particularly critical in scattering environments where inhomogeneous illumination degrades contrast.

SSIM is defined as

$$SSIM = \left[\frac{2\mu_x\mu_y + C_1}{\mu_x^2 + \mu_y^2 + C_1}\right]\left[\frac{2\sigma_{xy} + C_2}{\sigma_x^2 + \sigma_y^2 + C_2}\right] \tag{11}$$

where $\mu$ and $\sigma$ are the mean values and standard deviations, respectively, with $x$ and $y$ representing two different images. It is worth noticing that $\sigma_{xy}$ means the structural covariance of two images, which means the ground truth and imaging result in our context. The two small constants $C_1$ and $C_2$ are used in eq. (11) to avoid division by zero. SSIM highlights the structural difference of the obtained images and reference ground truth, which is more perceptually aligned with how humans judge image quality.

Spatial resolution, governed by the speckle correlation length, determines the minimum discernible detail, with smaller speckle sizes enhancing resolution but often at the cost of reduced CNR due to increased noise sensitivity. For instance, hybrid speckle patterns and adaptive algorithms (e.g., deep neural networks) balance these trade-offs, achieving high-resolution imaging[29]. Moreover, Li et al. [30] proposed an image quality assessment method, which effectively solves the problem of existing methods. It performs very well on different databases, leading the intersection of the imaging and processing, and acting as a solid reference for image assessment.

These metrics collectively guide the optimization of GI systems, ensuring robustness in applications ranging from underwater exploration to dynamic scattering media analysis.

**3. Multi-wavelength ghost imaging**

GI exhibits wavelength-dependent characteristics that derive from the availability of modulators, detectors, and propagation physics at each spectral band. In the visible and near-infrared, programmable spatial-light modulators and high-efficiency silicon cameras yield high-fidelity reconstructions and support advanced variants such as light-field GI [55, 56] and non-degenerate schemes that illuminate in the infrared while detecting in the visible [66, 67], thereby exploiting mature detector technology. Moving into the ultraviolet (200–400 nm), GI circumvents the poor quantum efficiency and high cost of UV camera arrays by pairing a 325 nm laser with a single-pixel detector; recent demonstrations add dark-field and quantitative modalities that highlight surface defects and phase edges at low cost. In the mid-infrared (3–5 μm), difference-frequency generation transfers high-bandwidth patterns from telecom wavelengths to the mid-IR idler, enabling single-pixel temporal GI that accesses molecular-fingerprint bands while leveraging off-the-shelf modulators [76]. At terahertz frequencies (0.1–10 THz), where fast focal-plane arrays remain scarce, GI's single-pixel architecture and nonlinear emitters such as spintronic sources deliver sub-diffraction imaging [89] and 3-D micro-volumetry [80], although water absorption restricts penetration depth. In the hard-X-ray and extreme-ultraviolet (XUV) regimes, GI decouples illumination dose from image formation: synchrotron and tabletop sources have achieved micron-scale, ultra-low-dose X-ray GI, while free-electron-laser experiments extend the concept to nanometer-resolution XUV imaging of radiation-sensitive samples [45]. Across all bands, the common thread is that GI leverages second-order correlations to image where array detectors are inefficient, thereby enabling modality-specific advantages—from high-speed visible reconstructions to chemical selective mid-IR sensing and low-damage X-ray microscopy.

### 3.1 Ghost imaging with particles

Particle-based, or matter-wave GI extends the second-order correlation paradigm from photons to massive quanta, harnessing their de Broglie waves to create images with high sensitivity and intrinsically non-destructive probe–sample interactions. Because spatial degrees of freedom are read out by a single-pixel bucket detector, the technique is particularly attractive whenever position-resolved detectors are unavailable, inefficient, or would impart unacceptable radiation dose.

**Atoms.** The first experimental realization was reported by Khakimov et al. (2016) [31], who produced correlated atom pairs by colliding a Bose–Einstein condensate (BEC) of metastable helium. The atoms, sharing a single quantum state, served as the analogue of entangled photons in optical GI and enabled micron-scale imaging of a test mask. A follow-up study [32] exploited high-order correlations among ultracold atoms to boost image visibility without sacrificing spatial resolution, underscoring the value of many-body quantum statistics for noise suppression. Matter-wave GI with neutral atoms therefore provides a platform for probing macroscopic quantum phenomena while mitigating sample damage, an aspect that is attractive for delicate biological specimens.

**Neutrons.** Neutron GI pushes the concept into a regime where penetrating power, rather than resolution, is paramount. Because thermal neutrons traverse high-Z metals yet highlight light elements such as hydrogen, they complement X-rays in materials science and nondestructive evaluation. Kingston et al. [33] implemented GI on a poly-energetic reactor beamline, integrating spatial resolution into instruments that traditionally deliver only bulk flux readings while simultaneously reducing the neutron dose. Subsequent work by He et al. [34] introduced a single-pixel neutron imager employing $Gd_2O_3$ masks patterned in Hadamard sequences, further demonstrating dose-efficient, compressive neutron imaging.

**Electrons.** Electron ghost imaging combines the sub-nanometer resolving power of electron optics with the dose reduction inherent to bucket detection. Li et al. [35] employed a 266 nm UV-pumped cathode to generate patterned electron beams driven by a digital micromirror device (DMD). Their proof-of-concept experiment showed that acquisition time and specimen damage can be reduced by orders of magnitude relative to conventional scanning or full-field electron microscopy, paving the way for high-throughput, low-dose studies of radiation-sensitive materials.

Collectively, these demonstrations establish matter-wave ghost imaging as a versatile framework that complements photon-based GI. By judiciously selecting the particle

species—atoms for quantum-statistical studies, neutrons for deep penetration, or electrons for ultra-high resolution—researchers can tailor the technique to a wide spectrum of scientific and technological applications while retaining the core advantages of low dose, high sensitivity, and simple detection geometry.

**3.2 Ghost imaging with X-rays and XUV**

To achieve lensless imaging, researchers utilize the electromagnetic properties of light and sophisticated computational algorithms to form images without traditional lenses. Techniques such as holography [36], ptychography [37], and coherent diffraction imaging [38] reconstruct high-resolution images from diffraction patterns. In these patterns, the low-frequency components, typically in the central region, exhibit the highest intensities, while the high-frequency components are weaker, necessitating a high photon flux for enhanced resolution and detailed image reconstruction. Despite their capabilities, these methods depend on multi-pixel array detectors like complementary metal-oxide-semiconductor (CMOS) and charge-coupled device (CCD) technologies, which are readily available in the visible range. However, in other spectral ranges such as extreme ultraviolet (XUV/EUV), optimized cameras are both expensive and scarce.

Compared with visible lights, X-rays have shorter wavelength and higher energy. Due to their short wavelength, X-rays can penetrate substances that light cannot. Different materials absorb X-rays to varying degrees based on their density and atomic number. This property allows X-rays to pass through soft tissues more easily than through bones or metals[39]-[41]. However, X-rays are also capable of ionizing atoms and molecules, which made illumination dose a critical problem in X-ray imaging. Ghost imaging techniques have the potential to avoid phototoxicity caused by high light dose.

At almost the same time in 2016, two groups realized X-ray ghost imaging independently, marking the promised potential of X-ray GI. Yu et al. [42] introduced a novel method for lensless Fourier-transform ghost imaging Fourier-transform ghost imaging using pseudothermal hard X-rays. This technique extends X-ray crystallography to nanocrystalline samples by measuring the second-order intensity correlation function of light, allowing the Fourier-transform diffraction pattern of a complex amplitude sample to be achieved at the Fresnel region. Concurrently, Pelliccia et al. [43] provided an experimental proof-of-concept for ghost imaging in the hard X-ray energy range. Their setup used a synchrotron beam split by a crystal in Laue diffraction geometry, with the necessary illumination patterns arising from natural speckles in the thermal hard X-ray source emitted by the synchrotron.

Following initial proof-of-principle experiments, subsequent research began to focus on reducing the cost and complexity of X-ray GI. In 2017, Schori and Shwartz [44] investigated the feasibility of performing X-ray GI using an incoherent, low-brightness X-ray tube source. Their setup utilized a Rigaku SmartLab X-ray diffraction system with a rotating copper anode operating at 8.05 keV. The X-ray beam was split using a highly oriented pyrolytic graphite crystal: one beam passed through the object and was collected by a single-pixel detector, while the other, which did not interact with the object, was directed to a multi-pixel detector.

In 2018, Zhang et al. [45] reported a major advancement by demonstrating ghost imaging using a compact tabletop X-ray source with significantly reduced radiation levels. In a follow-up study [46], they incorporated deep learning techniques to enhance image reconstruction. This work achieved high-resolution imaging (~10 μm), suitable for applications such as early-stage cancer detection, while using only 18.75% of the Nyquist sampling rate—thereby minimizing both radiation exposure and system cost. A key innovation was the integration of Hadamard-patterned illumination with a multi-level wavelet convolutional neural network, greatly improving reconstruction efficiency from sub-sampled data.

Moreover, researchers have explored unconventional optical components to overcome the scarcity of suitable X-ray optics. In 2023, Li et al. [47] integrated polycapillary optics into an X-ray GI system to enhance resolution. Their work demonstrated that these optics could reduce the average unit size of the illumination pattern from 166 μm to 55 μm, effectively tripling the imaging resolution. There has also been extensive discussion around enhancing the imaging performance of X-ray GI. In 2018, Schneider et al. [48] introduced a novel imaging technique using incoherently scattered light from a free-electron laser (FEL), demonstrating the feasibility of ghost imaging under challenging coherence conditions. In a related study, Pelliccia et al. [49] explored the use of synchrotron radiation to improve the practical applicability of XGI. Their work focused on advanced image reconstruction methods, characterization of the system's point-spread function, and strategies for reducing radiation dose—highlighting the broader potential of XGI in scientific and biomedical research.

Moving beyond planar imaging, Kingston et al. [50] extended ghost imaging into the third dimension with the development of **ghost tomography**. This approach involves illuminating an object with spatially random X-ray intensity patterns from multiple angles, while a single-pixel detector records the total transmitted intensity. This method enables volumetric reconstructions and significantly broadens the application space of GI.

In addition, ghost imaging has been extended into the extreme ultraviolet (XUV) regime. Kim et al. [51] demonstrated the feasibility of XUV ghost imaging using a free-electron laser. In their experiment, each FEL pulse was passed through a moving diffuser to produce a sequence of speckled, uncorrelated intensity patterns.

### 3.3 Ghost imaging in visible range

Experimental progress in GI within the visible spectrum has primarily advanced along two fronts: sophisticated illumination schemes and data-driven reconstruction techniques. In underwater environments, the short-wavelength end of the visible range—particularly violet and blue light (400–500 nm)—offers superior penetration due to reduced absorption and molecular scattering. Wang et al. [52] leveraged this spectral window using compressive computational ghost imaging (CGI) with Hadamard patterns and total variation priors, later enhancing edge recovery through wavelet-domain processing. Chen et al. [53] demonstrated that temporal CGI, combined with a low-bandwidth avalanche photodiode, can enable error-free optical data transmission across meter-scale underwater channels—highlighting GI's resilience for concurrent imaging and communication. Illumination engineering has also progressed significantly. Pseudo-Bessel ring [54] and Lorentz-beam [55] modulations have been employed to harness their self-healing properties, mitigating the effects of beam wander and intensity fading. These structured illumination strategies markedly enhance image contrast at greater depths.

Deep learning has recently revolutionized visible-band GI, dramatically enhancing performance under low sampling conditions. Wu et al. [56] introduced DAttNet, a deep residual attention network that lowers sampling ratios far below the Nyquist limit while preserving high fidelity reconstructions, and Li et al. [57] combined compressive sensing with a super-resolution CNN to detect centimeter-scale objects with a single photodiode. These data-driven models ingest low-dimension bucket signals and learn the inverse mapping to high-resolution scenes, outperforming handcrafted priors in both clarity and speed. Chen et al. [58] addressed target-irrelevant speckle interference in CGI with a self-supervised dual-network scheme, as shown in Fig. 3(a). The first network mines task-relevant cues directly from the raw speckle field, discarding redundant components, while the second refines the provisional reconstructions by isolating true object features from background clutter. Training is entirely self-supervised: both subnetworks are driven to make the predicted bucket-signal sequence converge to the experimentally measured sequence. Consequently, the pipeline dispenses with labelled datasets and shows strong cross-scene generalization in challenging underwater conditions; the reconstruction quality depends mainly on the bucket-signal "label," with minimal sensitivity to input variations. Li et al. [59] later introduced a block-wise CGI model that also uses 2-D random patterns as inputs. By decomposing the image into multi-scale sub-blocks and iteratively feeding its own reconstructions back into the network, the model learns fine details while progressively injecting prior knowledge, as shown in Fig. 3(b). Using the one-dimensional bucket signal as the sole supervisory cue further enhances robustness and transferability. These studies collectively demonstrate that substituting conventional CGI inputs with 2-D random patterns does not compromise reconstruction quality. Building on this insight, Geng et al. [60] proposed a multi-input, mutual-supervision framework in which a random-image branch (rich spatial priors) and a bucket-signal branch (high physical fidelity) co-train by supervising each other's outputs, as shown in Fig. 3(c). The complementary interplay of the two inputs markedly improves accuracy, robustness, and the correction of low-quality image artefacts.

To embed even richer object information in the supervisory signal, Lu et al. [61] captured one-dimensional intensity sequences under multiple polarization states and fused them as training labels in Fig. 3(d). Their multi-polarization CGI framework first records linear- and circular-polarization bucket signals in transmission or reflection, then analyses them in a multi-branch fusion network that exploits the statistical differences between polarization modes. During training, the fused multi-polarization sequence constrains parameter convergence, boosting generalization. Experiments confirmed high-quality reconstructions across a range of environments, including simulated, aerial, and strongly scattering underwater scenes.

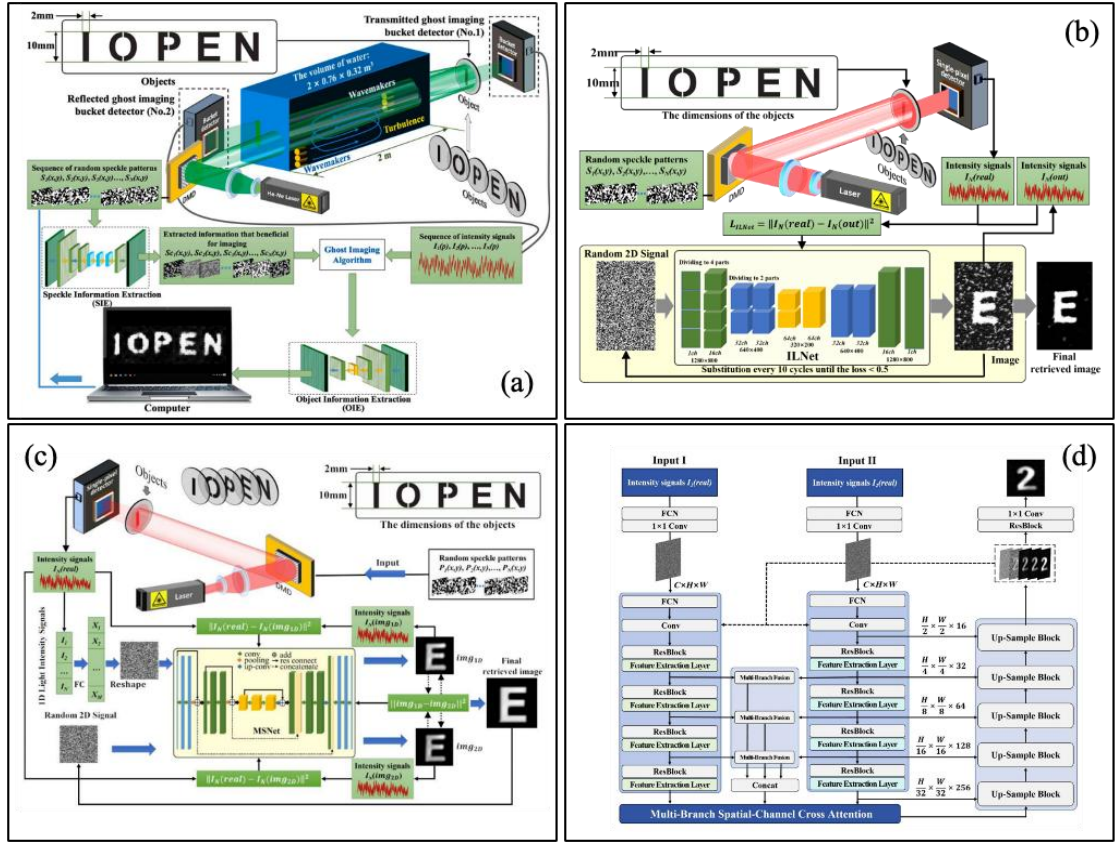

Fig. 3 Framework of supervised ghost imaging (a) Self-supervised feature extraction-based ghost imaging; (b) Part-based ghost imaging; (c) Multi-input mutual supervision ghost imaging; (d) Multi-polarization fusion network for ghost imaging.

Despite rapid progress, three key challenges continue to hinder the practical deployment of GI systems. First, the illumination hardware must operate at multi-kilohertz refresh rates while maintaining phase stability in corrosive, high-pressure environments—conditions that currently exceed the capabilities of most DMDs and SLMs. Second, achieving video-rate reconstructions will require hybrid algorithms that integrate wave-optics physics into lightweight neural architectures. These models must minimize the number of bucket measurements without compromising image quality or interpretability. Third, future GI systems must be self-optimizing: capable of continuously monitoring water clarity, turbulence, and ambient lighting, and dynamically adjusting the operating wavelength, mask ensemble, and exposure time to maximize the signal-to-noise ratio in real time.

### 3.4 Ghost imaging with Infrared light

Infra-red ghost imaging (IR-GI) inherits the single-pixel, fluctuation-correlation paradigm of its visible-band ancestor yet leverages the atmospheric windows at 0.8–2.5 μm, 3–5 μm and 8–12 μm to see through fog, smoke and dust, work at photon-starved fluxes, and avoid photo-damage—traits prized in covert surveillance, remote sensing, non-destructive inspection and biomedical diagnostics [62]-[65]. Radwell et al. [66] proved the concept by replacing the focal-plane array with a single InGaAs detector in a dual-band microscope that records visible and short-wave-IR images simultaneously, recovering micron-scale detail without cryogenic sensors. Synchronizing a high-speed digital-micromirror device with an efficient compressive algorithm, the same group pushed the technique to 10 Hz real-time video through smoke and tinted glass, delivering visible–SWIR footage with a single detector [67]. They then extended IR-GI into the mid-IR: a single-pixel camera tuned to the 1.65 μm methane line captured video-rate images of leaking plumes, an early demonstration of passive gas-leak monitoring with GI [68].

Near-infrared (NIR) ghost-imaging techniques have rapidly moved out of the laboratory and onto airborne and spectroscopic testbeds. Wang et al. [69] built a centimeter-resolution, three-dimensional ghost-imaging LiDAR that couples a sparsity-constrained inverse solver to a single time-resolved detector, enabling depth maps to be recorded from a moving aircraft. Broadband single-pixel hyperspectral cameras now classify concealed chemicals across 900–1700 nm at only milliwatt illumination levels by exploiting compressive sampling and bucket-signal inversion, as demonstrated by Gattinger et al. [70] in their SWIR single-pixel imager for chemical characterization. Additionally, ghost spectroscopy has incorporated quantum-correlated (and classical thermal) photon pairs. Janassek et al. [71] showed that wavelength-correlated photons provide molecular-fingerprint specificity while conferring intrinsic eavesdropping immunity, extending the ghost-imaging paradigm to secure spectral sensing.

Infra-red ghost imaging in the mid-IR (3–12 μm) has moved from laboratory curiosities to photon-starved, high-speed instruments that challenge the performance of conventional focal-plane arrays [72]-[75]. Temporal GI normally relies on ultrafast modulators and detectors, limiting its use in spectral bands where such hardware is scarce—particularly the mid-infrared (MIR). Wu et al. [76] overcome this barrier with a frequency-downconversion TGI scheme, random intensity patterns are imprinted on a near-IR signal and transferred to an MIR idler (3.2–4.3 μm) via difference-frequency generation in a nonlinear crystal. A slow MIR detector then captures the pattern-encoded waveform, and correlation processing reconstructs ultrafast temporal objects without scan or high-speed electronics. The concept is wavelength-agnostic and paves the way for pump-probe imaging and ultrafast dynamics studies in MIR, THz, and other bands that lack fast modulators or detectors.

Because DMD suffers severe diffraction losses beyond ≈ 3 μm, pattern delivery is migrating to electrically tuned graphene–metasurface modulators that steer centimetre-scale beams at gigahertz rates and promise megapixel counts unattainable with MEMS mirrors [77]. At even longer wavelengths, passive thermal ghost imaging reconstructs scenes using only self-emitted radiation, enabling non-contact temperature maps of hot machinery or living tissue in complete darkness and aligning naturally with applications where active light would disclose the sensor or perturb the sample[78]. Image quality now rivals that of cooled focal-plane arrays thanks to physics-aware deep networks that embed the GI forward model, achieve super-resolution without pre-training and remain interpretable across imaging conditions[79]. When combined with adaptive denoising, saliency-guided sampling and the coming convergence of ultralow-noise single-pixel detectors and gigahertz modulators, these learning-augmented solvers position mid-IR ghost imaging as a robust, cost-effective alternative for covert surveillance, remote chemical mapping and non-destructive inspection in environments where conventional optics falter.

Future, the convergence of ultralow-noise single-pixel detectors and room-temperature up-conversion, gigahertz-speed metasurface modulators that overcome DMD bandwidth and diffraction limits, and learning-augmented, physics-constrained solvers capable of real-time, saliency-aware reconstruction positions mid-IR ghost imaging for rapid translation into field systems.

**3.5 Ghost imaging with terahertz waves**

Terahertz ghost imaging (THz-GI) blends the non-ionizing, fabric-penetrating power of 0.1–10 THz radiation with correlation-based reconstruction, making it ideal for security screening, pharmaceutical QC and cultural-heritage diagnostics[80]-[84]. Because THz waves sail through textiles, polymers and paper yet carry molecular-fingerprint contrast, they can reveal hidden defects that visible or X-ray systems miss[85]-[91]. Practical deployment, however, is shaped by three hurdles. First, strong atmospheric water-vapor absorption. Second, the scarcity of fast, high-contrast spatial light modulators beyond the mid-IR. Third, the data burden of high-dimensional compressive inversion[92]-[96]. In particular, the tensor-based algorithm [97] is a significant technique to effectively process the high-dimensional data with retaining the original property of data.

Terahertz (THz) spectroscopy excels at probing a sample's chemical composition and material properties, yet high-resolution imaging at these wavelengths remains a challenge due to the scarcity of suitable detector arrays. Olivieri et al. [98] address this gap theoretically by proposing a nonlinear ghost-imaging scheme that surpasses conventional single-pixel methods where multi-pixel detectors are unavailable. The concept fuses nonlinear THz generation with time-resolved field sampling—capabilities provided by modern THz time-domain spectroscopy—and targets hyperspectral imaging of semitransparent samples that introduce significant propagation delays. Full-wave, time-resolved acquisition enables accurate spatiotemporal reconstruction of complex inhomogeneous structures, offering a clear pathway toward high-resolution THz imaging beyond current hardware limits. Chan et al. [99] replaced pixelated detector arrays with a single photoconductive antenna and random printed-circuit-board masks, inaugurating THz single-pixel, compressive imaging. Shen et al. [100] soon swapped static masks for a rotating random disk, pushing acquisition toward video rate and underscoring the rate-limiting role of modulation hardware. Optical excitation of carriers in crystalline Si delivered the first dynamically reconfigurable THz spatial light modulator, enabling fully programmable masks and laying the foundation for single-pixel imaging without mechanical parts [101]. Watts et al. [102] then introduced electrically tunable metamaterial modulators that switch in sub-millisecond times with centimetre apertures, achieving high-throughput, compressive THz imaging on a single detector. Parallel routes exploited optically pumped germanium masks [103] and, more recently, spintronic emitter arrays that embed the pattern directly in the THz generation process, slashing optical losses and alignment complexity while delivering deep-sub-λ near-field resolution [104].

The path forward is being forged on four key fronts. First, reconfigurable metamaterials and two-dimensional materials are enabling gigahertz-speed, centimeter-scale modulators. Second, advances in low-noise detectors are mitigating the penalties of atmospheric absorption. Third, wavefront shaping combined with iterative reconstruction techniques is pushing spatial resolution toward the diffraction limit. Finally, physics-informed deep neural networks are drastically reducing the number of required measurements while preserving interpretability.

Together, these developments are converging toward compact, turnkey THz ghost imaging systems capable of operating in humid environments, resolving fine structural details, and delivering real-time, non-invasive inspection across industrial, medical, and security applications.

## 4. Conclusion

GI has evolved from its quantum-entangled origins into a versatile, correlation-based framework spanning the extreme-ultraviolet to terahertz regimes—and even matter-wave analogues with atoms, neutrons, and electrons. By decoupling illumination from detection, GI replaces costly or unavailable focal-plane arrays with single-pixel detectors and computational reconstruction, enabling high-fidelity imaging where conventional cameras fall short.

In multi-wavelength contexts, this architecture supports dose-efficient X-ray and XUV microscopy, chemically selective mid-infrared sensing, sub-wavelength terahertz inspection, and ultra-low-noise visible/near-IR 3D lidar—each tailored via custom modulators. In scattering media, GI's second-order intensity correlations suppress phase-randomizing noise, yielding clear images through turbid water, fog, biological tissue, and turbulent atmosphere

Recent breakthroughs stem from three converging trends. First, reconfigurable modulators now achieve gigahertz speeds and broad spectral tunability, reducing acquisition times from minutes to real-time video rates. Second, physics-informed compressed sensing and deep neural networks reconstruct high-resolution scenes from heavily under-sampled data, slashing measurement counts by more than an order of magnitude while preserving interpretability. Third, adaptive, wavelength-agile platforms incorporate environmental feedback to dynamically optimize signal-to-noise ratios in changing conditions.

**Abbreviations**

GI: Ghost imaging

XUV/EUV: extreme-ultraviolet

THz: terahertz

CGI: computational GI

DGI: differential GI

CSGI: compressive sensing GI

DLGI: deep learning ghost imaging

CMOS: Complementary Metal Oxide Semiconductor

CCD: Charge-coupled Device

SLM: spatial light modulator

DMD: digital micromirror device

CS: compressive sensing

RMSE: root mean squared error

PSNR: peak signal-to-noise ratio

CNR: contrast-to-noise ratio

SSIM: structural similarity

MSE: mean squared error

BEC: Bose–Einstein condensate

FEL: free-electron laser

XGI: X-ray GI

CNN: Convolutional Neural Network

IR-GI: Infra-red ghost imaging

SWIR: short-wave-IR

NIR: Near-infrared

LiDAR: Light detection and ranging

MIR: mid-infrared

MEMS: Micro-Electro-Mechanical Systems

**Data Availability Declaration**

All data analyzed during this study are derived from published sources, which are cited in the manuscript. No new data were generated.

**Competing Interest Declaration**

The authors declare that they have no competing interests.

**Author Contribution Declaration**

C.S. conceived and designed the review. T.T. performed the literature search and manuscript writing. S.O. participated in discussion and manuscript reviewing. C.S.

supervised the project and contributed to the interpretation of findings. All authors contributed to writing and revising the manuscript and approved the final version.

**Acknowledgements**

This work was supported by the Cluster of Excellence "Balance of the Microverse" (EXC 2051 – Project-ID 390713860) funded by the Deutsche Forschungsgemeinschaft (DFG, German Research Foundation).

**Code availability**

Not applicable.

**Funding**

This work was supported by the Cluster of Excellence "Balance of the Microverse" (EXC 2051 – Project-ID 390713860) funded by the Deutsche Forschungsgemeinschaft (DFG, German Research Foundation).